\documentclass[num-refs]{wiley-article}

\usepackage{siunitx}
\usepackage{todonotes}

\usepackage{tikz}

\usepackage{xcolor}     % for colour
\usepackage{ntheorem}   % for theorem-like environments
\usepackage{mdframed}   % for framing

\theoremstyle{break}
\theoremheaderfont{\bfseries}

\newmdtheoremenv[%
linecolor=gray,leftmargin=60,%
rightmargin=40,
backgroundcolor=gray!40,%
innertopmargin=0pt,%
ntheorem]{myprop}{Recommendation}

\usetikzlibrary{mindmap}

\papertype{Original Article}
\paperfield{Journal Section}

\title{Towards Continuous Compounding Effects and Agile Practices in Educational Experimentation}

\author[1]{Luis M. Vaquero, PhD}
\author[1]{Niall Twomey, PhD}
\author[1]{Miguel Patricio Dias, PhD}
\author[1]{Massimo Camplani, PhD}
\author[1]{Robert Hardman, PhD}

\affil[1]{KidsLoop Ltd, London, E16BX, United Kingdom}
\corraddress{Luis M. Vaquero PhD, KidsLoop Ltd, London, E16BX, United Kingdom}
\corremail{luis.vaquero@kidsloop.live}

\runningauthor{Vaquero et al.}

\begin{document}

\begin{frontmatter}
\maketitle

\begin{abstract}

Randomised control trials are currently the definitive gold standard approach for formal educational experiments. Although conclusions from these experiments are highly credible, their relatively slow experimentation rate, high expense and rigid framework can be seen to limit scope on: 1. \textit{metrics}: automation of the consistent rigorous computation of hundreds of metrics for every experiment; 2. \textit{concurrency}: fast automated releases of hundreds of concurrent experiments daily; and 3. \textit{safeguards}: safety net tests and ramping up/rolling back treatments quickly to minimise negative impact. This paper defines a framework for categorising different experimental processes, and places a particular emphasis on technology readiness. 

On the basis of our analysis, our thesis is that the next generation of education technology successes will be heralded by recognising the context of experiments and collectively embracing the full set of processes that are at hand: from rapid ideation and prototyping produced in small scale experiments on the one hand, to influencing recommendations of best teaching practices with large-scale and technology-enabled online A/B testing on the other. A key benefit of the latter is that the running costs tend towards zero (leading to `free experimentation'). This offers low-risk opportunities to explore and drive value though well-planned lasting campaigns that iterate quickly at a large scale. Importantly, because these experimental platforms are so adaptable, the cumulative effect of the experimental campaign delivers compounding value exponentially over time even if each individual experiment delivers a small effect.


\end{abstract}

\end{frontmatter}

\section{Introduction}
\label{sec:intro}

\begin{epigraph}{Dr Seuss in ``Horton Hears a Who!''}
Don't give up! I believe in you all. A person's a person, \textbf{no matter how small}! And you very small persons will not have to die. If you make yourselves heard! So come on, now, and TRY!
\end{epigraph}

\begin{figure}[bt]
\centering
\includegraphics[width=10cm]{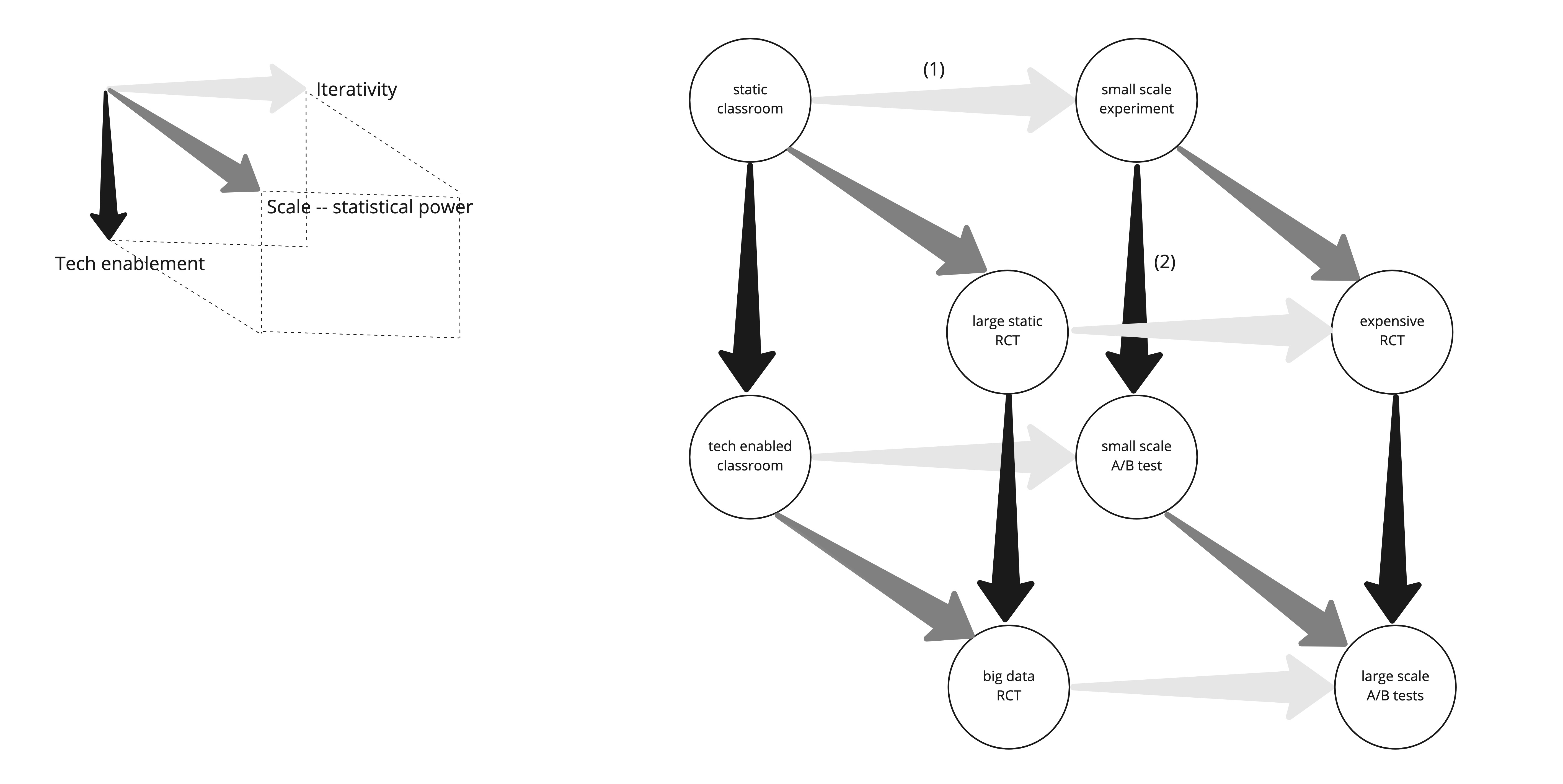}
\caption{Classroom Experimentation Cube: starting from the top left corner of the cube one can explore where the addition of iterative experimentation (such as the one happening in the classroom), large number of participants (scale - like in large RCTs) and advanced technology support would lead to in terms of type of experiment being produced.}
\label{fig:cube}
\end{figure}

The fundamental topic of consideration in this paper is experimentation in educational settings. We consider three primary axes listed here as a framework for categorising and comparing experiments: 1. iterativity, 2. scale, and 3. technological readiness. These axes are shown visually in Figure \ref{fig:cube}, and the set of nodes results in eight distinct experimental types. The most foundational and common node is the traditional static classroom that are often void of direct experimentation, but will follow recommended practices for teaching. Each axis leaving this root node `adds' new capabilities to the learning environment, and the remaining nodes in the cube are the cases where one, two or three new capabilities. 

Concretely, it is well-known that most educational experiments occur in small and localised environment as part of daily teaching practise. This experimentation process can be distinguished from the `static classroom' by moving along the `iterative' axis (edge number (1) in Figure \ref{fig:cube}). It is possible to augment this testing environment further by introducing technology that can, for example, capture data and coordinate several teachers/classes in the same year or even several schools in a district. This shifts the experimental categorisation down edge the `technology enablement' axis (edge (2) in Figure \ref{fig:cube}). In both cases it is important that best practices for experimental are followed so that they do not incur potentially harmful and costly initiatives on children \cite{Torgerson2001}. 
Using best practice guides from the medical literature as an example, experiments are expected to be double-blind and effect and sample size calculations should be conducted \textit{a priori}. This is particularly critical when experimenting may lead to the dissemination of public guidance for best education practice. However, it is easy to see the challenges in guaranteeing these requirements in several cases in the cube: teacher-led experimentation cannot be double-blind, and sample size is often determined by the class size and cannot be negotiated in small experiments. In other words, although these are valuable experiments, they will likely need to be validated by a larger-scale experiment before wide-scale adoption can be considered.

Randomised Controlled Trials (RCTs) are the widely accepted standard of experimentation and hypotheses testing across disciplines. Owing to their long and successful legacy in many contexts (particularly medical), they have garnered deep trust, offering many safeguards against poor treatments and outcomes, and statistical guarantees that successes will be detected. It is little surprise, therefore, that RCTs are the \textit{de facto} model for educational experimentation. 
The majority of recent successes in educational experimentation have aligned with this. However, even when rigorous evaluations are performed using randomised experiments, they often lack the iterative nature of low scale classroom led experimentation (see ``large static RCT'' and ``big data RCT'' in Figure \ref{fig:cube}). Also, the resulting effect sizes are much smaller than previous research in the same substantive area \cite{Dawson2018}. 

This means lots of resources are wasted and key decisions result in irrelevant impact on the education of the learners. Achieving the required rigour has traditionally been linked to slow, costly processes that require enormous infrastructure and funding schemes (see ``expensive RCT'' in Figure \ref{fig:cube}). Government funding and coordination is often required to ensure the implementation of educational RCTs \cite{Pontoppidan2018}. To establish good RCT practice in the U.S., the Institute of Education Sciences (IES) was introduced by the Education Science Reform Act of 2002. The mission of the IES is to build a body of rigorous evidence to inform education policy and practice \cite{Hedges2018} (\url{http://ies.ed.gov}). Similar governance processes have been instigated in Europe, with the Education Endowment Foundation (EEF) in the U.K. overseeing over 100 education RCTs in a very short period of time in \cite{Connolly2018}. 
RCTs have several weaknesses, however, including high expense, recruitment challenges, sample size calculation, implementation and testing. 

On the other hand, cheap and reliable experiments are routinely conducted by nearly every technology company. These randomised \textit{online} controlled experiments, often called A/B tests, are used to drive decisions about how to improve their products. A/B tests and RCTs both rely on the same family of experimental methods, and have been tried and tested millions of times for over a century. Although RCTs and A/B tests have many similarities (particularly on their theoretical foundations), there are significant differences on how they are applied in general. RCTs tend to be expensive, have longer follow up times, and be extra rigorous in terms of planning the analysis, resulting in a very slow and bureaucratic process overall. On the other hand, A/B tests are inexpensive, fast, and encourage quick tests to prune down non productive interventions early. 

A key distinction between the two that is worth mentioning explicitly here is the consequence and cost of failed experiments. In many online trials, up to 70\% of the promising ideas fail to improve the core outcome metrics, and these rates approach 80–90\% for well-optimised systems \cite{Kohavi2013}. \textit{A priori}, therefore, it is expected that approximately 9 out of 10 experiments will fail, similar to the teacher-led experiments in the classroom. Despite these poor odds, online testing repeatedly and consistently deliver enormous value because they are designed to be resilient to failure. Resilience is achieved with rapid experimental pivots, testing small incremental changes fast and continuously, technology has enabled methodologically consistent, statistically rigorous, cost effective RCTs at an incredible scale. A/B tests are delivered through/with the assistance of online devices such as phones, tablets or computers. 

With the migration of many aspects of education to digital devices, the opportunity to learn, experiment, and improve education quickly and at scale is at hand. Adopting the techniques from technology companies can bring the potential to dramatically alter the scale and pace of how trials are conducted in education. Rather than seeing educational interventions as single-shot experiments seeking large changes, embracing technology-like A/B test means adopting a long chain of (potentially) small effects that compound to significant long-term benefits. The combination of the rapid experimentation of teacher-led small scale experiments with the rigour of large scale RCTs has been enabled by the availability of new technologies at scale (see ``large scale A/B test'' in Figure \ref{fig:cube}). By following the best experimental practice and performing experiments at scale, highly credible results should be acquired. In turn, these may be appraised by a committee of educational experts, augment statistical power through conducting meta-analyses and become a key component for making public recommendations of best education practice, ultimately affecting all nodes in the experimentation cube.

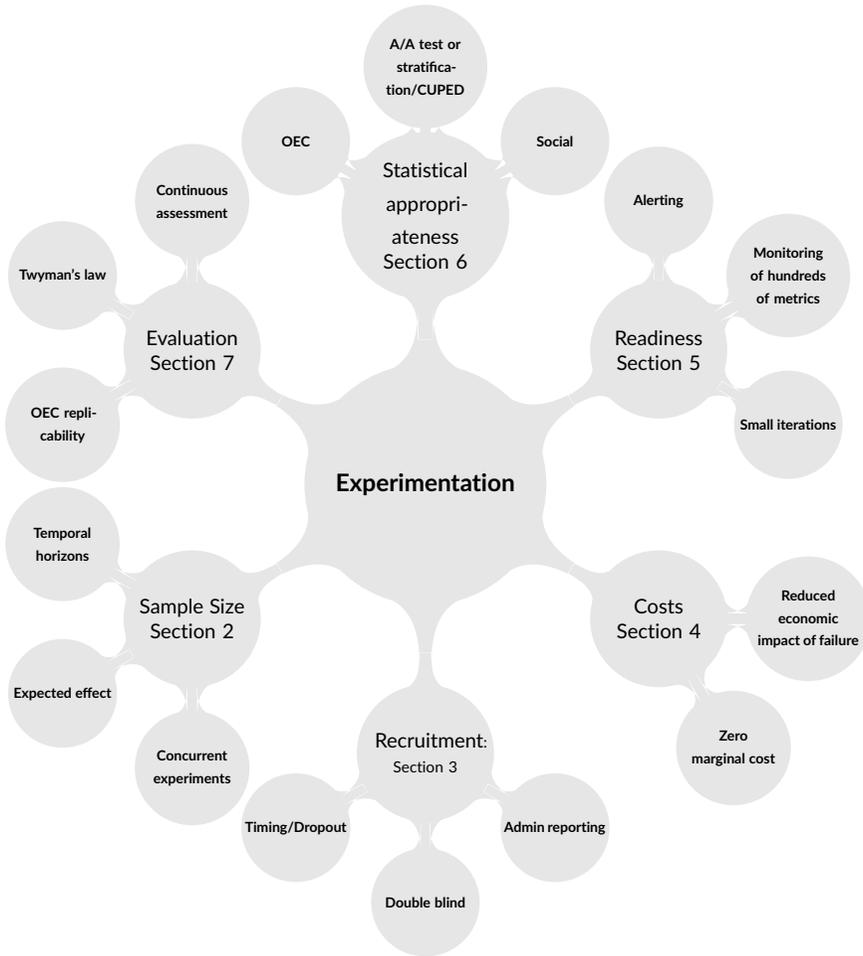
\begin{figure}
    \centering
    \resizebox {0.815\textwidth} {!} {
    \begin{tikzpicture}[
        mindmap,
        every node/.style=concept,
        concept color=black!10,
        grow cyclic,
        level 1/.append style={level distance=4.5cm,sibling angle=60},
        level 2/.append style={level distance=2.5cm,sibling angle=60}
        ]
      \node [root concept] {\textbf{\Large Experimentation}} % root
        child [concept] { node {\large {Sample Size} \\Section \ref{sec:samplesize}}
          child { node [concept,text width=6em,text centered] {\textbf{Temporal horizons}}} 
          child { node [concept,text width=6em,text centered] {\textbf{Expected effect} }}
          child { node [concept,text width=6em,text centered] {\textbf{Concurrent experiments} }}
        }
        child [concept,text width=6em,text centered] { node {{\large{Recruitment}}:\\Section \ref{sec:recr}}
          child { node [concept,text width=6em,text centered] {\textbf{Timing/Dropout}  }}
          child { node [concept,text width=6em,text centered] {\textbf{Double blind}}}
          child { node [concept,text width=6em,text centered] {\textbf{Admin reporting}}}
        }
        child [concept] { node {{\large{Costs} \\Section \ref{sec:cost} }}
          child { node [concept,text width=6em,text centered] {\textbf{Zero marginal cost} } }
          child { node [concept,text width=6em,text centered] {\textbf{Reduced economic impact of failure} } }
        }
        child [concept] { node {\large {Readiness} \\Section \ref{sec:ready}}
          child { node [concept,text width=6em,text centered] {\textbf{Small iterations}}} 
          child { node [concept,text width=6em,text centered] {\textbf{Monitoring of hundreds of metrics} }}
          child { node [concept,text width=6em,text centered] {\textbf{Alerting} }}
        }
        child [concept] { node {\large {Statistical appropriateness} \\Section \ref{sec:appr}}
          child { node [concept,text width=6em,text centered] {\textbf{Social}}} 
          child { node [concept,text width=6em,text centered] {\textbf{A/A test or stratification/CUPED} }}
          child { node [concept,text width=6em,text centered] {\textbf{OEC} }}
        }
        child [concept] { node {\large {Evaluation} \\Section \ref{sec:eval}}
          child { node [concept,text width=6em,text centered] {\textbf{Continuous assessment}}} 
          child { node [concept,text width=6em,text centered] {\textbf{Twyman’s law} }}
          child { node [concept,text width=6em,text centered] {\textbf{OEC replicability} }}
        };
    \end{tikzpicture}
    }
    \caption{This figure illustrates the main challenges for education RCTs that are covered by this paper and lists the relevant sections where we show how A/B testing can dramatically improve the situation.}. An Overall Evaluation Criterion (OEC) is a (usually composite) quantitative measure of the experiment’s objective,  when a single key performance indicator is deemed insufficient for the evaluation of the outcome of an online controlled experiment and using individual separate evaluation of metrics is not desirable for some reason. Controlled-experiment Using Pre-Existing Data (CUPED) is a variance reduction technique that uses previous data to control for natural variation in an experiment’s main metric. By removing natural variation, we can run statistical tests that require a smaller sample size. \textit{Temporal horizons} represent the duration of the treatment and the follow up period after the treatment stops/changes. 
    \label{fig:outline}
\end{figure}

Figure \ref{fig:outline} shows how education RCTs could improve by embracing some of the techniques and elements that are common in an A/B testing context. We start by presenting the main current challenges in using RCTs in education: sample size (Section \ref{sec:samplesize}), recruitment (Section \ref{sec:recr}), cost (Section \ref{sec:cost}), readiness (Section \ref{sec:ready}), and rigour (Sections \ref{sec:appr} and \ref{sec:eval}). All of these sections present technical and cultural changes induced by the potential introduction of A/B tests that can help address/mitigate current limitations. Finally, Section \ref{sec:remain} discusses how a hybrid between RCTs and A/B tests can solve most of these challenges so that the education community may consider whether and how similar designs can be applied in a focused fashion or at massive scale.

\section{Calculating the Appropriate Sample Size}
\label{sec:samplesize}

\subsection{Technical Change}
As platforms scale up, more experiments are required and experimenters need to spot incrementally smaller effects after the first initial big wins in not-optimised platforms are realised. Given that the relationship between the detectable effect and the required umber of users for a fixed statistical power is quadratic \cite{Simon2009}, four times more users are required to detect effects half as large. 

This limitation has led to relatively small (75\% including fewer than 1000 learners) or excessively long (many months) studies \cite{Connolly2018}. Indeed, a major difference between A/B testing in technology and educational RCTs that dramatically affects sizing is their time horizons: learning takes time and progress and impact can take years to realise. 

As time goes by in a study, the sample size also increases and the 95\% confidence  interval (which is proportional\footnote{Assuming Treatment and Control have the  same  size and  same  population  variance and samples are independent and identically distributed} to the expression in Equation \ref{eq:confidence}) tends to decrease (smaller error). This assumes the coefficient of variation  (CV),  the  ratio  of the standard  deviation to  the  mean, does not change for the duration of an experiment. However, empirical  data  shows that CV  does change over  time especially for long-lasting experiments. 

\begin{equation}
\frac{CV}{\sqrt{sample size}}
\label{eq:confidence}
\end{equation}

For this assumption to hold, experiments need to be shortened and, in order to keep statistical power high, many more users will be needed \cite{Kohavi2020OnlineRC}: in practice, \textbf{experimenters will need to expose each user to multiple experiments at the same time}. Tech companies design campaigns of experiments that are shorter in duration where the immutability of CV can be guaranteed. 

Consider the following example that illustrates population requirements in single vs. multi-experiment enrolment for students. Let's assume that experimental design specifies that 3,000 students are required to understand significance of an experiment, and let us further assume that we wish to run 50 experiments simultaneously. If this experimental campaign assigns students to only one experiment, then 150,000 unique students must be recruited. On the other hand, if students can simultaneously be in 5 experiments on average (this may be reasonable if each experiment targets different aspects of the product), then we reduce our recruitment burden dramatically by a factor of 5. As shown in Section \ref{sec:appr} classroom settings impose some additional considerations when determining the right test to estimate required size/power.

\begin{figure}[bt]
\centering
\includegraphics[width=10cm]{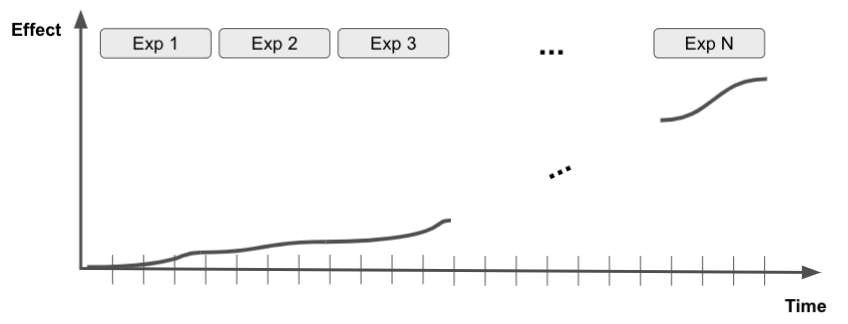}
\caption{Experiment campaign compounds a large effect as the result of a long succession of short experiments with a modest individual effect each of them.}
\label{fig:compound}
\end{figure}

\begin{myprop}
Education RCTs may need to be to be realised as a planned campaign of shorter experiments aiming to reveal smaller effect sizes over large populations that can be subject to many experiments in parallel. The timeframe for compounding effects is large and aligned with learning as a long life activity.
\end{myprop}

Figure \ref{fig:compound} shows how a campaign of short-lived experiments delivering small effects can build up over long periods of time to a massive population, resulting in potentially big impacts for the learners and society. 

There are many examples of large tech companies using overlapped/concurrent experiments \cite{Kohavi2013,Tang2010,Xu2015}. To preserve the robustness and rigour of the experiments, they had to develop mechanisms to prevent interactions (e.g., by declaring constraints or parameters being modified, the system will guarantee disjoint users to those experiments) and overnight tests on all pairs of experiments to detect interactions. 

Cross-talk tests would have to look at effects across several variables such as country, language, gender to see if a tests is exerting noise on a subgroup enrolled in a different test. Despite the theoretical risk of experiment cross-talk, practical results have shown that interactions in online platforms tend to be relatively rare and more often represent bugs than true statistical interactions \cite{Kohavi2013}. Small multi-variate tests that evaluate the impact of multiple variables that could interact can be run to check for suspected cross-talks between experiments.

\begin{myprop}
Online education platforms would have to preserve as much contextual information as possible and elaborate tests to check things such as agonistic/antagonistic effects between subjects that the student is taking. 
\end{myprop}

\section{Recruiting Schools and Retaining them in Trials}
\label{sec:recr}

Recruiting schools can be a challenge as RCTs normally contribute with extra load to their already busy days. \cite{Dawson2018} suggest that timing can be critical: recruiting schools in the summer (national assessments, school trips) tends to result in poor engagement and higher attrition rates. Attrition may affect the statistical tests if some students drop out of the control/treatment group. Schools often welcome being part of the intervention group and the attrition rate in the control group tends to be much higher \cite{Dawson2018}.

In addition, teachers need to be trained and supported to ensure appropriate conditions are met. The UK's EEF recommends recruiting periods of at least 3 months or, ideally, 5-6 months \cite{Dawson2018}, which is a significant fraction of the work considering  that 70\% of the educational studies last less than one term \cite{Connolly2018}.

\subsection{Technical Change}

Online platforms enable dynamic recruitment of users into experiments. They check the conditions for inclusion/exclusion are met and potential confounders suggested to the experimenter. 

This dynamic enrollment ability means the user is automatically and inadvertently enrolled in a double-blind test and they remain agnostic of the fact that they may be part of one or more experiments. 

Another element to preserve engagement will be delivering insight about progress to key stakeholders (e.g. school principals). 

\begin{myprop}
A/B testing platforms enable a seamless recruitment that can be done dynamically (as enrolment conditions are met) and requires little (if any) training/effort  for the teachers as ``treatment`` is delivered in a blind manner for students and teachers alike, with frequent (live) results available in an aggregated form.
\end{myprop}

\section{Ensuring Cost Effectiveness}
\label{sec:cost}

In addition to the sizing/timing and enrollment practicalities of RCTs, they also come with notorious financial considerations. The cost of enrolling students/schools, obtaining consent, setting up the experiment, and analysing the results is minimised through the usage of online experimentation. 

A cost-efficient RCT in 1970 would be \$10000 (the equivalent of \$65000 in 2021's money) \cite{Welch1972}. Other authors quantify the cost as a function of the cost to measure an expected size.  Interventions that cost less than \textsterling80 per pupil per 0.1 shift in the standard deviation of the effect are considered very good value for money, less than \textsterling170 per pupil per 0.1 standard deviations is good value for money (all of these are 2018 British Sterling Pound) \cite{Higgins2015,Dawson2018}.

\subsection{Technical Change}

The marginal cost or running a new RCT in an online platform tends to null once it is built and the cost is amortised across the millions of learners in the platform and the built-in automation of tests and process such as recruitment results in negligible monetary impact.   

As the experimentation platform matures, running and analysing experiments becomes self-service.

This means that users without coding skills can design and setup an experiment with assistance to automatically recruit and randomise schools or individual children in a double-blind manner. Also, an online experimentation platform includes guidance/automation of data processing to ensure statistical rigour in the obtained results (see Section \ref{sec:appr} below).

\subsection{Cultural Change}

This affordability makes some of the classic considerations about size of the effects and cost less relevant  \cite{Dawson2018}. 

\begin{myprop}
Near zero cost per new user means new elements to be taken into consideration, such as the need to constantly (in near real-time) check for key metrics ensuring progress and protection from harmful effects for the learners. 
\end{myprop}
\section{Ensuring Interventions are Ready for Trial}
\label{sec:ready}

\subsection{Technical Change}

According to \cite{Dawson2018}, some of the interventions funded by the EEF were simply not ready for trial. The responsibility for this problem is shared by: the developers, for seeking funding for something that has not previously been adequately tested for evidence that it could work. 

Lack of readiness in education trials resulted in time and money waste. However, when the experimentation platform is no longer limiting the number of experiments, organisations embrace a ‘test everything with controlled experiments’ mentality. The limiting factor to innovation now becomes the ability to generate ideas and develop the code for them adding automated precautions not to make harmful interventions for the learners.

In A/B tests, experiments are set and defined using a web browser with 1000s of metrics computed for each experiment with the first read on the experiment impact in minutes for critical metrics.

\begin{myprop}
Calculating hundreds or thousands of metrics automatically for every experiment in near real-time means that if certain key metrics degrade beyond acceptable limits, the experiment is terminated automatically. If after several hours no key metric degrades, the experiment auto-ramps up to a higher percentage of users and at multiple locations.

Also, fast turn around of experiment results can help to abort obviously bad experiments by triggering alerts. 
\end{myprop}

Alerting  on any statistically  significant negative metric  changes  will  lead  to  an  unacceptable  number  of false  alerts.  To avoid this \cite{Kohavi2012} recommend: 

\begin{itemize}
    \item Changes need to be not only statistically significant, but also large enough in absolute magnitude  to  have  meaningful  user, data quality,  or  business  impact.
    \item Corrections  for  multiple  testing. The  O’Brien and Fleming procedure suggests using  lower  p-values  early  on  and  to increase them over time, so that  an  experiment  that  if the results are extreme we gain the benefit of stopping early.
    \item Different  magnitudes  of changes for  different  metrics  are categorised   in   specific  severity   levels. The   most   severe changes result  in  automatic  shutdown  of  an  experiment  but less severe changes will result in emails sent to the owner of the experiment and a central experimentation team.
\end{itemize}

\subsection{Cultural Change}

The availability of a cost efficient platform to run experiments resulted in release cycles going from 6 months to monthly, to weekly, to daily, to multiple times a day  \cite{Kohavi2020OnlineRC}. 

Indeed, testing ideas before implementation resulted 70\% of the promising ideas failing to improve the metrics they were designed to improve (even worse for well-optimised systems, where failures were in the range of 80–90\% \cite{Kohavi2013}). 

Cheap, reproducible, controlled experiments encourages more exploration of ideas that may not be highly prioritised \textit{a priori}, but are easy to code and evaluate. There is no strong correlation between the effort to code an idea and its value \cite{Kohavi2020OnlineRC}. For example, small changes to Google’s colour scheme, were worth over \$200M annually.

\begin{myprop}
Having an online experimentation platform that supports the amalgamation and archival of different forms of evidence (e.g. causal inference methods) and be useful to stage RCTs and help design RCT campaigns better \cite{Morrison2021}.
\end{myprop}
\section{Choosing and Delivering Appropriate Testing}
\label{sec:appr}

\subsection{Technical Change}

A key assumption for A/B tests to deliver good results is the: ``Stable Unit Treatment Value Assumption'',  which  states  that  the  behavior  of  each sample in the experiment depends only on their own treatment and not on the treatments of others \cite{Rubin74}. It is easy to see how a user who is served better search results is more likely to click, and that behavior is entirely independent of others using the  same  search  engine. 

However, such  assumption  does  not always  hold  in  experiments  run  on education where a learner's behaviour is highly likely to be influenced by her neighbourhood. For example, if a video  lesson  is confusing, learners will share that confusion leading to lower usage and/or more discussion about it for learners in both, treatment and control groups. 
  
In an A/B experiment,  this  implies  that  if the  treatment  has  a  significant impact  on  a  user,  the  effect  would  propagate  over  to  his/her  neighbourhood, regardless  whether  his/her  neighbors  are  in  treatment  or control. This is similar to social networks such as Facebook,  Twitter and  LinkedIn in which many features are likely to  have network effects \cite{Kohavi2020OnlineRC}.

Thus, a learner's  network connections need to be taken into account when sampling them into treatment and  control by: 1) partitioning the learners into clusters and 2) treating each cluster as a unit for randomisation so all users in one cluster have the same experiment assignment \cite{Gui2015,Kohavi2020OnlineRC}. One big drawback with this clustering approach is its vulnerability to carryover effects, where the same users who were impacted by the first  experiment  are  being  used  for  the  follow-on  experiment although simple A/A  tests  can  be  run  to  check  for carryover effects triggering cluster membership randomisation at the expense of a temporary loss of experiment capacity \cite{Kohavi2012}. This is much more important in education where interventions are expected to have a positive long lasting effect. 

\begin{myprop}
Online platforms make it easy to deliver experiments where treatment and control users receive identical experience (known as A/A tests). In this setting, any statistically significant differences for each metric should happen at about 5\% (when using a p value cutoff of 0.05) can be attributed to chance or bias \cite{Kohavi2020OnlineRC} such as hidden dependencies or carry-over (residual) effects with previous experiments affecting subsequent experiments on a subgroup of users.
\end{myprop}

Moreover, education RCTs tend to rely on clusters (schools or classrooms) as a basic randomisation unit, which can make some experimental designs non-standard \cite{Spybrook2020}. These authors highlight the need to take cluster-level and individual level into account to answer the ``under what conditions'' and ``for whom'' questions, making calculations more complicated than for standard RCTs or A/B tests.

Randomisation may not work all the time like when users are keener in treatment than in control just by chance (not due to the Hawthorne effect or a willingness to be treated). Stratification can be expensive to implement efficiently during the sampling phase, so many online platforms start by dynamically checking the equilibrium of key metrics using historical data. If they are unbalanced, there is a re-randomisation process that uses a different hash ID. Seeds for the hash function can be used to determine which one leads to a difference that is not statistically significant \cite{Kohavi2012}, using post-stratification or CUPED \cite{Deng2013}.

\begin{myprop}
Education RCTs delivered as A/B tests will require rigorous testing for:
\begin{itemize}
\item cross talk when users can be enrolled in multiple experiments
\item individual metric variation and composite OEC effect
\item stratification effects and re-randomisation to balance different strata
\item ability to re-run tests for reproducibility 
\end{itemize}
\end{myprop}

\subsection{Cultural Change}

Appropriate testing means the company builds trust in the process of incrementally gathering evidence about what works and does not work. As a result, organisational goals tend to shift from ‘deliver feature X by date Y’ to ‘improve the Overall Evaluation Criterion (OEC) by x\% over the next year’. After a few initial easy wins, most ideas are not as good as we believe \cite{Kohavi2014} and many interventions end up impacting on secondary outcomes, without major long term progress on the main OEC. This makes the need to iterate on small incremental gains that build up in the same way compound interest does. 

\begin{myprop}
Rather than focusing just on the average treatment effect, having a score card of key metrics helps to unveil key segments of users where the treatment effect is different than the average. This enables hypotheses to develop and experiments start to target segments of users. In contrast to typically underpowered subgroup analyses in education trials, these are highly powered with a surplus of users that make the segments big enough for reliable statistical analyses.
\end{myprop}
\section{Robust Implementation and Process Evaluation}
\label{sec:eval}

\subsection{Technical Change}

With so many experiments running in parallel for every learner, education practitioners will be concerned about lack of trustworthiness and false positive results. Online platforms perform multiple tests to identify scenarios that would indicate a problem \cite{Fabijan2019}. 

For instance, if the experiment design is for equally sized control and treatment then deviations in the actual ratio of users in an experiment likely indicate a problem and the experiment needs to be tagged as invalid. There are simple tests that help us identify numerous issues in experiments, many of which invoke Twyman’s law (``Any figure that looks interesting or different is usually wrong'') \cite{Kohavi2020}.

Long spacing between assessments (like in Norway and Sweden) makes assessment results less well suited for the short interventions often evaluated within RCTs, as post-assessments either have to be made long after the intervention ended or pre-assessments are not current enough \cite{Pontoppidan2018}. 

\begin{myprop}
Online platforms enable a nearly continuous evaluation of the results of the assessments. 
\end{myprop}

\subsection{Cultural Change}

An OEC is a (usually composite) quantitative measure of goal of an experiment (also known as Response or Dependent Variable, Outcome Variable, Evaluation metric, Performance metric). OECs are used when a single key performance indicator (KPI) is not sufficient for the evaluation of the outcome of an online controlled experiment and a weighted combination of KPIs is used instead.

Agreement on the OEC becomes critically important, with hundreds to thousands of metrics changing at the same time. The challenge is then coming up with a small set of key metrics, ideally a single OEC, to help make trade off decisions. 

A good OEC should not be short term (e.g. clicks, time to completion) as it should capture the organisational long-term objectives (e.g. predicted lifetime value, compound acceleration of learning rate, content engagement, and repeat visits), but must be based on metrics that are measurable in short-term experiments. 

A classic example in online technological companies is sessions per user metric (also units purchased, revenue, profit, expected lifetime value, or some weighted combination of these): if users are coming more often, it is usually a strong sign that the treatment is useful. This metric may not be ideal in education as increased screen time is not always welcome by parents and teachers. The definition of appropriate metrics including not just engagement, but also how productive a learning session was seems much more relevant for learning applications. 

\begin{myprop}
In online experimentation platforms, team performance is measured on the sum of treatment effects of a single key metric. All experiments with a positive result are rerun (hence replicability is key), so that the replication run determines the actual credit the team gets. The replication effect is unbiased, while the first run may have found an exaggerated effect \cite{Gelman2014}.
\end{myprop}

This technical element also helps mitigate the replication crisis that has plagued other scientific fields and is likely to be a problem in education science \cite{Hedges2018b}. 

\section{Discussion and Conclusion}
\label{sec:remain}

Technology is evolving, and unless we can overcome the inertia of today's education process many opportunities will be missed.

\begin{table}[]
\caption{Comparison between RCTs and A/B Tests}
\label{table:comparison}
\begin{tabular}{p{0.2\textwidth}p{0.3\textwidth}p{0.4\textwidth}}
\multicolumn{1}{c}{\textbf{}} & \multicolumn{1}{c}{\textbf{RCT}} & \multicolumn{1}{c}{\textbf{Online A/B Tests}} \\
\textit{Platform}             &   Offline / Human-led         &    Online / automated         \\
\textit{Inclusion and exclusion criteria} &    Complex, controlling for confounding factors, mediators and comorbidities    &    Simple rules based on product interaction  \\
\textit{Recruitment}          & Slow, manual communications for ethics and informed consent  & Blanket consent obtained with end user license agreement agreed when using the product      \\                        
\textit{Awareness}             & Participants are well aware they are part of a study and may know if they are part of the control or the treatment group (not blind)  &  Participants tend to be unaware they are part of a study and also whether they are part of control or treatment    \\                        
\textit{Concurrency}  &   Participants are involved with at most one RCT  & Participants are recruited for many studies in parallel     \\                        
\textit{Metrics}              &    Limited in number and  in measurements times   &  Hundreds or thousands of metrics measured continuously \\  
\textit{Iteration and rectification}&    Slow one-offs; difficult to rectify aspect of the study  & Campaign of rapid tests planned ahead; the iterative approach allows for changes in a succession of tests     \\  
\textit{Duration} & Months or years and firm termination dates (relying on funding) &  Days or weeks for individual tests; months for campaigns with no firm termination (premature ending or run in perpetuity)    \\  
\textit{Dynamic treatment}  & static, well-defined treatment applied uniformly over the experiment) & `treatment' may be a machine learning model or algorithm that can be re-fitted with new data and adapt to the context \\  
\textit{Consequence of failure}  & High - significant monetary and time costs associated & Low - `fail fast' early termination and better time efficiency  \\  
\textit{Publication} & External - aiming to generalise conclusions   & Internal - results are only valid for the product at hand \\  
\end{tabular}
\end{table}

Table~\ref{table:comparison} shows the main differences between RCTs and A/B tests. These differences are mainly derived from the very different setting in which RCTs and A/B tests are applied. RCTs tend to be applied in academic environments where process and rigour are the most important elements, event at the cost of significantly higher additional expenses and timelines. Hence, RCT participants are very carefully selected and are aware they are part of an RCT. They can only be involved in that RCT where a handful of metrics and mediators are measured over a long period of time (months to years). A/B tests diminish the cost and accelerate iteration speed and do not directly feed recommendations back to the public domain. 

\subsection{Accelerate with Small Increments}

Hybrid testing that incorporates the strengths of A/B tests and RCTs should deliver rapid, trustworthy and continuous experimentation in the educational domain and should lead to actionable recommendations for the public education. The current generation of RCTs in education fall short of these ambitions principally due to prohibitively slow turnaround time. We believe that a paradigm shift of experimentation process is overdue, and that the next generation of experimentation in education will be heralded by incorporating modern experimentation practices. 

Rather than designing for large size long lived experiments, education RCTs would have to shift to a sequence of short campaigns with smaller effects (and large sample sizes) that compound over time. This is compatible with current thinking in the domain that argues that small size effects requiring large sample sizes that are pedagogically sound and easy to implement, have an important impact \cite{Dawson2018}. Failing fast and moving on to the next idea seems to be better \cite{Kohavi2012}.

\subsection{Continuous Evaluation}

The speed of analysis required by A/B testing is calling for a change in how evaluation is performed. Assessments tend to be formative (aimed at informing the teacher to steer the class and target the materials) or summative (more formal student knowledge tests). The former are frequent and often done at the discretion of each teacher and not captured by online platforms; the latter are scarce and often mandated by an authority. Online platforms should allow for a more frequent of assessing the learners without them knowing they are actually being assessed.

Many schools and organisations tend to focus on the assessment of learning outcomes as a key element to evaluate performance. Other metrics, such as productive engagement (minimising time on an online platform but maximising learning) or enjoyable focused time will have to be defined to help educational A/B tests delver their value into the education domain. 

\subsection{Beyond Current Limits}

Interventions targeting hard-to-reach groups or school structures remain difficult in education \cite{Dawson2018}. Online platforms can mitigate this challenge by enabling the experimentation on subsets of the population that would be difficult to identify or pool together otherwise.  A/B testing platforms enable the dynamic enrollment of pupils upon meeting some conditions (e.g. struggling with a piece of content of a given kind for more than X weeks in a row).

\citet{Hedges2018b} highlights the generalisability of education research beyond the context where it was tested on. While this may remain a challenge, scaling content utilisation to millions of learners and makes the chances of generalisation beyond the online platform the tests were performed on more likely. A/B testing and the ability to create an archive of experiments that can be used for later meta analysis can be helpful. Indeed, comparing the ``causal web`` between any two experiments may prove a useful tool for generalisation \cite{Morrison2021}.

The possibility of determining information and knowledge propagation has been explored at a single course level in observational studies \cite{Vaquero2013,  Vaquero2019}. A/B testing in online educational platforms may enable the application of epidemiology-like techniques in education. 

\subsection{Diminishing Returns}

RCTs are known to present issues when adapting to increasingly complex interventions \cite{Hedges2018b}. The same is true for A/B tests: the first tests tend to result huge successes, but with each fruitful intervention the percentage of improvement becomes smaller. However, a succession of  cheap, rapid and simple interventions, albeit small, cause a two-fold compound effect: 1) over time on each single learner and 2) over the whole population when applied at scale.

\subsection{Linking Back to Theory}
RCTs have been criticised for just explaining the why and not the how, making it hard to find mechanistic explanations on how these interventions cause their effects \cite{Hedges2018b}. A/B tests are no different. This situation is also similar to how medical research happens: an RCT discovers an unexpected effect and this triggers a wave of more basic research operating at a molecular level trying to dissect the explanation for the observed evidence. This is the basis of inductive thinking.

In addition, well thought campaigns of A/B tests can be used as tools to gather the evidence to prove or refute a theory in deductive thinking scenarios.

The classic questioning about whether researchers know more about what constitutes evidence than teachers, but less than policy makers \cite[p. 21]{Morrison2021} is diluted by the utilisation of an experimentation platform that reduces technical and operational barriers and can be used by all of them alike, ensuring comparable rigour and standards. 

\subsection{Not Universal Applicability}

RCTs are not suited to answering all kinds of questions, and there are some things to which schools are just not willing to be randomised, such as ideological questions, financial incentives for teachers that not all schools could follow, school timetables, etc. 

A/B tests tend to be imperceptible by teachers and learners, blind to them and are embedded into common practice, therefore reducing the impact of the Hawthorne effect and helping to make evidence more robust. \cite{Morrison2021} (p. 21) highlights how educational studies should be asking ``under what circumstances does this work``. The wide access to millions of learners and teachers and iterative nature of A/B tests can help determine specific conditions for certain interventions to work. Moreover, it helps organisations build a historical memory of things that have not worked in their specific context. A/B tests platforms could learn about how to do meta-analyses to extract insight from A/B tests performed in different campaigns or settings.

A/B testing platforms tend to include tools and support for forms of evidence other than RCTs (e.g. observational and causal inference), which can help with how RCTs are prepared and screened and enable triangulation of data from many directions \cite{Morrison2021}.

\section*{Acknowledgements}
We'd like to thank Sandy Suh, Rath Bala, Mona Nainie, and PJ Williams for their useful feedback to prior versions of this manuscript.

\section*{Conflict of interest}
The authors are with an EdTech company running A/B tests to make content adapt to the needs of the learner.

\printendnotes

% Submissions are not required to reflect the precise reference formatting of the journal (use of italics, bold etc.), however it is important that all key elements of each reference are included.
\bibliography{sample}

% \begin{biography}[example-image-1x1]{A.~One}
% Please check with the journal's author guidelines whether author biographies are required. They are usually only included for review-type articles, and typically require photos and brief biographies (up to 75 words) for each author.
% \bigskip
% \bigskip
% \end{biography}

% \graphicalabstract{example-image-1x1}{Please check the journal's author guildines for whether a graphical abstract, key points, new findings, or other items are required for display in the Table of Contents.}

\end{document}